\documentclass[aps,prb,twocolumn]{revtex4-1}
\usepackage[a4paper,top=2.00cm,bottom=2.00cm,left=2.00cm,right=2.00cm]{geometry}
\usepackage{amsmath}
\usepackage{amssymb}
\usepackage{latexsym}
\usepackage{enumerate}
\usepackage{tgbonum}
\usepackage[utf8]{inputenc}
\fontfamily{cmr}\selectfont
\usepackage{graphicx}
\DeclareGraphicsExtensions{.eps, .pdf, .jpg, .png}
\usepackage[usenames,dvipsnames]{color}
\usepackage{xcolor}
\usepackage[colorlinks=true,urlcolor=blue,linkcolor=blue,citecolor=blue]{hyperref}
\usepackage{bookmath}
\usepackage{mymatrices}
\def\omegasq{\omega_{\text{sq}}}
\def\omegarsq{\omega_{\text{rsq}}}
\begin{document}
\title{
\vspace*{-20mm}\hspace*{-50mm}
{\tiny 
Published in Phys.Rev.Lett. 47, 7, 530 (1981) , J. A. Sauls
\href{https://doi.org/10.1103/PhysRevLett.47.530}{[https://doi.org/10.1103/PhysRevLett.47.530]}
}
\\
{$\ns$}\vspace*{15mm}
\\
{Soliton Propagation of Zero Sound in Superfluid \Heb}
} 
\author{J. A. Sauls}
\affiliation{Joseph Henry Laboratories,
         Princeton University, Princeton, NJ 08544}
%\email{sauls@lsu.edu}
\date{May 29 1981}
\begin{abstract}
Nonlinear equations for zero sound interacting with the B-phase collective modes are derived by treating \Heb\ as a medium of two-level “molecules” (Cooper pairs), The nonlinear equations reduce to known results from kinetic theory in the linear response limit. Soliton solutions to the nonlinear equations are discussed in context with the experiments of Polturak et al.
\end{abstract}
%---------------------------------------------------------------------------------
\maketitle
%---------------------------------------------------------------------------------

The condensate of Cooper pairs in superfluid \He\ has many similarities to a system of Bose-condensed diatomic molecules. Leggett~\cite{leg80} has reviewed these similarities for the equilibrium phases of \He-A and -B. In addition to ``molecular'' properties of the ground-state pairs, there are numerous collective modes of the order parameter that correspond to excited bound-pair states. Several of these modes have been observed in sound propagation experiments in both superfluid phases.~\cite{whe78,gia80,mas80} 

Recently Polturak et al.~\cite{pol81a} reported solitonlike propagation of zero sound in \He-B when the zero-sound frequency is near a collective-mode resonance. These authors point out that the nonlinear effects they observe are similar to those of nonlinear optical systems.~\cite{mcc69} This article makes that analogy explicit.

Besides the well-known squashing (sq) mode,~\cite{wol78a} with excitation energy $\hbar\omegasq\simeq 1.5\Delta(T)$, a new collective mode has been observed in B-phase sound propagation experiments at $\hbar\omegarsq\simeq 1.0\Delta(T)$.~\cite{gia80,mas80} The identification of this mode has recently been clarified.~\cite{koc81,ave80,sau81} Koch and W\"olfle~\cite{koc81} showed that the \emph{real} squashing (rsq) mode, with total angular momentum $J=2$, couples weakly to zero sound because of particle-hole asymmetry. The fivefold Zeeman splitting observed by Avenel, Varoquaux, and Ebisawa~\cite{ave80} confirms the identification with the rsq mode. Recent explanations for the temperature dependence and anomalous shift of the rsq mode away from $\hbar\omegarsq=\left(\nicefrac{8}{5}\right)^{\text{1/2}}\Delta(T)$ further support this identification.~\cite{sau81}

For present purposes the linear theory of zero sound in \He-B presented in Ref.~\onlinecite{wol78b} is the most efficient. The continuity equations for the mass density and momentum density are
\ber
\partial_t\delta\rho + \vec\nabla\cdot\vec{g} = 0\,, \qquad\quad && 
\label{eq-continuity-density}
\\
\partial_t\vec{g} + \vec\nabla P + \vec\nabla\cdot\dyadic{\pi} = 0 \,, &&
\label{eq-continuity-current}
\eer
where $\delta\rho$ and $\vec{g}$ are the density fluctuation and momentum density, and $P = c^2\delta\rho$ is the diagonal part of the momentum-stress tensor which plays the role of the pressure in the collisionless limit; $c$ is the zero-sound velocity. 

The coupling of sound to the order-parameter modes enters through the stress tensor, $\dyadic\pi$. The dissipative part of $\dyadic\pi$ from quasiparticle collisions is neglected. In this case
\be
\pi_{ij} = g'\partial_t\,d_{ij}^{(-)} + \kappa'\,d_{ij}^{(+)}
\,,
\label{eq-stress_tensor}
\ee
where $d_{ij}^{(-)}$ ($d_{ij}^{(+)}$) is proportional to the imaginary (real) part of the $J=2$ time-dependent order parameter, $g'=\nicefrac{1}{3}n\lambda\hbar$, $\kappa''=\nicefrac{8}{5} n\lambda\Delta\eta$, $n$ is the number density, $\Delta$ is the BCS gap, $\lambda$ is the dimensionless coupling constant of order one, and $\eta\simeq (\Delta/2\epsilon_F)\,\times \ln(0.1\epsilon_F/\Delta) < 10^{-2}$ is the particle-hole asymmetry factor. For longitudinal sound with $\vec{k}=k\,\hat{z}$, only the $J_z=0$ component of $d_{ij}^{(\pm)}$ couples to sound. Consequently, the density fluctuation satisfies
\be
\left(\partial_t^2 - c^2\,\partial_z^2\right)\delta\rho 
= 
g'\,\partial_z^2\,\partial_t\delta^{(-)}
+
\kappa'\,\partial_z^2\,\delta^{(+)}
\,,
\label{eq-wave_equation_sq+rsq_sources}
\ee
and the $J_z=0$ amplitudes, $\delta^{(\pm)} = \hat{z}_i\hat{z}_j\,d_{ij}^{(\pm)}$, satisfy
\ber
\left(\partial_t^2 + \omegasq^2\right)\delta^{(-)} &=& g\,\partial_t\delta\rho
\,,
\label{eq-sqmode}
\\
\left(\partial_t^2 + \omegarsq^2\right)\delta^{(+)} &=& \kappa\,\partial_t^2\delta\rho
\,,
\label{eq-rsqmode}
\eer
with $g=1/\hbar m N(0)$, $\kappa=2\eta/\Delta m N(0)$, and $N(0)$ the density of states at the Fermi surface.~\cite{wol78b}~\footnote{The coupling to the rsq mode can be obtained from Ref.~\onlinecite{koc81}.} The coupling between the density and order parameter fluctuations, in the linear theory, leads to resonant absorption and anomalous dispersion at the squashing-mode frequencies. 

Nonlinear collisionless kinetic equations describing sound propagation in superfluid \He\ have not been derived from a first-principles approach. Here nonlinear equations for zero sound are obtained by treating \He-B as a medium of ``molecules'' (Cooper pairs) in which transitions between states are induced by interaction with the density fluctuations of zero sound. A two-level description is suggested by the identification of the collective modes as excited bound-pair states,~\footnote{This identification is clearly demonstrated for the analogous modes in superconductors by P.C. Martin in \emph{Superconductivity}, edited by R. D. Parks (Dekker, New York, 1969), Vol. 1, p. 384.} and by the discrete spectrum of order-parameter collective modes, which is known to be anharmonic from detailed calculations~\cite{sau81} and from the observation of saturation of the sound attenuation.~\cite{pol81a} The ingredients of this theory are: (1) the equations of motion for the amplitudes of the two-level system, (2) the identification of the nonlinear amplitude that couples the density fluctuation and order parameter, and (3) the requirement that the nonlinear equations reduce to the known linearized equations for \He-B. Here we neglect the coupling to the sq mode since the nonlinear effects reported in Ref.~\onlinecite{pol81a} occur near resonance with the rsq mode; the splitting between modes ($\sim\nicefrac{1}{2}\Delta$) is large compared with the spectral width of typical sound pulses. The dynamics of the two-level system is assumed to be described by the equations
\be
i\partial_t\begin{pmatrix}u\cr v\end{pmatrix}
=
\begin{pmatrix} 
\omegarsq	&	i\kappa\,\partial_t\delta\rho 
\cr
-i\kappa\,\partial_t\delta\rho & 0
\end{pmatrix}
\begin{pmatrix}u\cr v\end{pmatrix}
\,,
\label{eq-two-level_dynamics}
\ee
where $u$ and $v$ are the amplitudes for the excited (rsq) and ground Balian-Werthamer (BW) states. For longitudinal sound in the B-phase it is possible to write the equations in terms of scalar amplitudes because $\delta\rho$ couples only to the $J_z=0$ component of the $J=2$ excited state and to the $J=0$ BW state. The density fluctuation which induces transitions between these two states satisfies the linear wave equation [Eq.~\eqref{eq-wave_equation_sq+rsq_sources} without the coupling to the sq mode],\footnote{Modifications to the left side of Eq.~\eqref{eq-wave_equation_sq+rsq_sources} appear unnecessary. The nonlinear behavior reported in Ref.~\onlinecite{pol81a} is observed only when the sound wave frequency is near a collective mode resonance.} \emph{but} with $\delta^{(+)}$ given by
\be
\delta^{(+)} = \nicefrac{1}{2}\left(uv^* + vu^*\right)
\,.
\label{eq-delta_vs_u+v}
\ee

This identification of the nonlinear amplitude that couples to the density fluctuation in Eq.~\eqref{eq-wave_equation_sq+rsq_sources} is required by gauge invariance and correspondence with the linearized kinetic theory. In particular, the coupling of the real squashing mode to the density fluctuation is proportional to the spin trace of $\left(\delta\hat\Delta^{\dag}\,\hat\Delta + \hat\Delta^{\dag}\,\delta\hat\Delta\right)$, where $\delta\hat\Delta$ is the time-dependent off-diagonal mean field and $\hat\Delta$ is the ground state off-diagonal mean field.~\cite{koc81} These mean fields are proportional to the pair amplitudes: $u\propto\delta\hat\Delta$ and $v\propto\hat\Delta$.~\footnote{To be precise, $\delta\hat\Delta=(i\sigma^{\mu}\sigma^{2})\,d_{\mu\nu}\,\hat{p}^{\nu}$ with $d_{\mu\nu}=u\Delta\times(\hat{z}_{\mu}\hat{z}_{\nu}-\nicefrac{1}{2}\delta_{\mu\nu})$, and $\hat\Delta=(i\sigma^{\mu}\sigma^{2})\,A_{\mu\nu}\,\hat{p}^{\nu}$, with $A_{\mu\nu}=v\Delta\,\delta_{\mu\nu}$. In the linear theory $v=1$; also note that the phase oscillations of the BW state have already been eliminated in favor of the density fluctuations.} 
The linear theory assumes small perturbations to the ground state so $v=1$; in this case the coupling is proportional to $\Re(u)$. The amplitude $\delta^{(+)}$ in Eq.~\eqref{eq-delta_vs_u+v} preserves the gauge invariant form for the coupling required by the microscopic theory, and allows for depletion of the ground state described by a time dependent $v$. The equations for $\delta^{(+)}$ (from here on $\delta$) and the other relevant bilinear products, $R=\left(uu^*-vv^*\right)$ and $\gamma=\nicefrac{1}{2i}\left(uv^*-vu^*\right)$, are
\ber
\partial_t\,R &=& 4\,\kappa\,\left(\partial_t\delta\rho\right)\,\delta
\,,
\label{eq-two-level_dynamics-R}
\\
\partial_t\,\delta &=& \omegarsq\,\gamma - \kappa\left(\partial_t\delta\rho\right)\,R
\,,
\label{eq-two-level_dynamics-delta}
\\
\partial_t\gamma &=& -\omegarsq\,\delta
\,.
\label{eq-two-level_dynamics-gamma}
\eer
%
%---------------------------------------------------------------------------------

Equations~\eqref{eq-wave_equation_sq+rsq_sources},~\eqref{eq-two-level_dynamics-R},~\eqref{eq-two-level_dynamics-delta}, and~\eqref{eq-two-level_dynamics-gamma} form the basic system of nonlinear equations for zero sound interacting with the rsq mode. When linearized about the equilibrium state with $v=1$, these equations reduce to Eqs.~\eqref{eq-wave_equation_sq+rsq_sources} and~\eqref{eq-rsqmode}, known results from linearized kinetic theory. Finally note that any amplitude dependence of the coupling constant $\kappa'$ is expected to be weak since $\lambda$ measures the fraction of single-particle excitations.  

In their paper on self-induced transparency, McCall and Hahn~\cite{mcc69} analyzed the propagation of light pulses in a medium of two-level atoms. At high power levels radiation pulses may propagate as solitons (at resonance with the two-level me- dium) with low-energy loss. The low attenuation that occurs for large amplitude pulses (self-induced transparency), even when low amplitude pulses are extinguished, results from stimulated emission of radiation back into the trailing edge of the pulse. Many of the characteristic features of soliton propagation have recently been observed in zero-sound experiments in \He-B. In particular, changes in pulse shape and velocity with the power level, and pulse scattering without distortion are reported.~\cite{pol81a}

%---------------------------- Figure 1 ----------------- 
\begin{figure}[t!]
\includegraphics[width=\columnwidth]{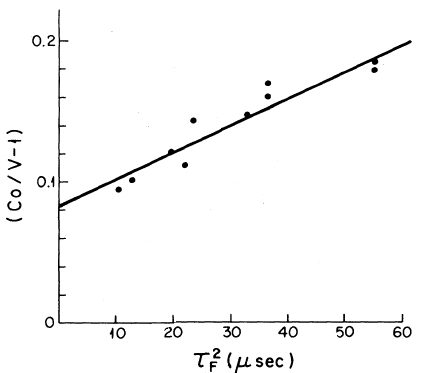}
\caption{Best fit of the pulse velocity data of Ref.~\onlinecite{pol81a} in the region between the onset of nonlinearity and the first step. The slope is $1.9\times 10^{-3}$ and the intercept is $0.083$. Note that $c_0$ is the unperturbed zero-sound velocity.}
\label{fig-solition_velocity}
\end{figure}
%----------------------------------------------------------------

These nonlinear effects are contained in the self-induced-transparency equations for zero sound interacting with the rsq mode. The equations are obtained by considering the evolution of sound pulses of frequency $\omega$ and wave vector $k=\omega/c$ which are modulated by a slowly varying envelope and phase shift. In particular assume
\ber
\delta\rho(z,t) &=& A(z,t)\,\cos\zeta
\,,
\label{eq-density_pulse}
\\
\delta(z,t) &=& D(z,t)\,\cos\zeta + E(z,t)\,\sin\zeta
\,,
\label{eq-rsq_pulse}
\eer
with $\zeta(z,t)=kz -\omega t + \Phi(z,t)$. $A$, $\Phi$, $D$, and $E$ are slowly varying functions on the scales $\omega^{-1}$ and $k^{-1}$. Since $R(z,t)$ measures the relative population of the two states, we expect $R$ to vary on the scale of the envelope function, provided the lifetime of the state is comparable to or greater than the pulse width. The self-induced transparency equations are obtained by neglecting higher-order derivatives of the slowly varying functions as well as higher harmonics of the fundamental frequency; also equations for $D$ and $E$ are obtained by first eliminating $\gamma$ from Eq.~\eqref{eq-two-level_dynamics-delta}. The resulting equations are
\ber
(\partial_t+c\partial_z)A = -\frac{\kappa'}{2c^2}\,\left[2c\partial_z D + \omega E\right]\,, &&
\label{eq-envelope-A}
\\
A\,(\partial_t+c\partial_z)\Phi = +\frac{\kappa'}{2c^2}\,\left[2c\partial_z E - \omega D\right]\,, &&
\label{eq-envelope-Phi}
\\
\partial_t D = \left[\Delta\omega - \partial_t\Phi-\nicefrac{3}{2}\kappa^2 A^2\,\omega\right]E\,, \qquad &&
\nonumber
\\
              - \kappa^2 AD\,\partial_t A + \,\kappa\,R\,\partial_t A\,, \quad && 
\label{eq-envelope-D}
\\
\partial_t E = \left[\partial_t\Phi - \Delta\omega + \nicefrac{1}{2}\kappa^2 A^2\,\omega\right] D\,, \qquad &&
\nonumber
\\
              + \kappa^2 AE\,\partial_t A -\,\nicefrac{1}{2} \,\kappa\,R\,\omega\,A\,, \quad && 
\label{eq-envelope-E}
\\
\partial_t R = 2\kappa\left[D\,\partial_t A + E\omega A\right]\,, \qquad\qquad\,\,\,\,\, &&
\label{eq-envelope-R}
\eer
%
%---------------------------------------------------------------------------------
%
where $\Delta\omega = \omega-\omegarsq$ is the deviation from exact resonance.

These equations simplify considerably for $\Phi$ constant. After rescaling, $\tilde{A}=\kappa\,\omega\,A$, $\tilde{E}=2E$, and $\alpha =\kappa\kappa'\,\omega^2/4c^2$, the consistent set of equations for the case of exact resonance is
\ber
(\partial_t+c\partial_z)\,\tilde{A} = -\alpha\,\tilde{E}
\,, &&
\label{eq-envelope-Atilde}
\\
\partial_t\,\tilde{E} = - \tilde{A}\,R
\,, \quad\qquad &&
\label{eq-envelope-Etilde}
\\
\partial_t\,R = \tilde{A}\,\tilde{E}
\,. \quad\qquad\,\,\,\,&&
\label{eq-envelope-R2}
\eer
%
%---------------------------------------------------------------------------------

These equations simply further with the transformation~\cite{lam71,cau73} $\tilde{A}=\partial_t\,\sigma$, $\tilde{E} = \sin\sigma$, and $R = -\cos\sigma$; in which case
\be
(\partial_t^2 + c\,\partial_z\partial_t)\,\sigma = -\,\alpha\,\sin\sigma
\,.
\label{eq-nonlinear_sigma}
\ee

Traveling pulse solutions, $\sigma(z,t) = \sigma(z-vt)$, satisfy
\be
\sigma^{''} - \alpha_v\,\sin\sigma = 0
\,,
\label{eq-nonlinear_sigma-pulse}
\ee
where $\alpha_v = \alpha/v(c-v)$. The standard one-solition solution, or ``$2\pi$-pulse'',
\be
\tilde{A}(z-vt) = 2\,\alpha_v^{\text{1/2}}\,v\,\sech\left[\alpha_v^{\text{1/2}}\,(z-vt)\right]
\,,
\label{eq-soliton_solution}
\ee
has a time integrated area of $2\pi$. Defining $\tau_p$, a measure of the pulse width in time, by 
$[2\alpha_v^{\text{1/2}}\,v]\,\tau_p = 2\pi$ ($\tau_p$ is related to the full width at half maximum 
by $\tau_F=0.84\tau_p$), then the pulse velocity and $\tau_p$ are related by
\be
\frac{v}{c} = \frac{1}{1 + \alpha\,(\tau_p/\pi)^2}
\,.
\label{eq-soliton_velocity}
\ee

The pulse velocity can be calculated from known coupling constants $\kappa$ and $\kappa'$, the sound velocity, and sound frequency. Measurements of group velocity dispersion by Calder et al.~\cite{cal80} give an experimentally determined value of the particle-hole asymmetry factor,~\cite{koc81} $\eta_{\text{exp}}=4.1\times 10^{-3}$. By using $\alpha=\nicefrac{4}{15}\,\lambda\,(\eta\omega v_F/c)^2$, $c/v_F=11$ for $P=24$ bar,~\cite{whe75} $\omega/2\pi=100$ MHz, and $\lambda\simeq 1$, then $\alpha\,(\tau_p/\pi)^2 = 2.1\times 10^{-3}\,\tau_F^2$, where $\tau_F$ is measured in $\mu$-sec. The pulse velocity varies from $v/c=0.98$ for $\tau_F=3\,\mu$-sec to $v/c=0.90$ for $\tau_F=7\,\mu$-sec, in qualitative agreement with the velocities reported in Ref.~\onlinecite{pol81a}.

Pulse compression and velocity increase with increasing amplitude, characteristics of solitons, are also reported in Ref.~\onlinecite{pol81a}. However, the two ``steps'' observed in the velocity versus amplitude are not characteristic of a single $2\pi$ pulse, but they may be due to interactions with additional solitons. Under the assumption that the data in Fig.~3 of Ref.~\onlinecite{pol81a}, following the onset of nonlinear behavior and up to the first step, correspond to the velocity of a single $2\pi$ pulse, then $(c/v) -1$ plotted against $\tau_F^2$ should be a straight line through the origin. The slope obtained from a least-squares fit (Fig.~\ref{fig-solition_velocity}) is $1.9\times 10^{-3}$ which compares with the theoretical value of $2.1\times 10^{-3}$. The finite intercept suggests that the sound velocity $c$ in Eq.~\eqref{eq-soliton_velocity} be interpreted as the zero-sound velocity modified by the small dispersion from the neighboring sq mode. With this interpretation, the intercept implies a dispersion correction to the unperturbed sound velocity $c_0$, of $c/c_0=0.92$, which agrees with the measurements by Calder et al.~\cite{cal80} for the sq mode dispersion at $\omega=\omegarsq$.

Damping of the modes has been neglected. The width of the rsq mode can be related to the normal-state quasiparticle lifetime, $\tau_{\text{N}}\,T^2=0.3\,\mu$sec\,mK$^2$, with use of the result of Koch and Wolfe.~\cite{koc81} For $T/T_c=0.4$ the rsq mode lifetime is $\gamma_{\text{rsq}}^{-1}\simeq 30\,\mu$-sec, which is considerably longer than typical pulse widths. Consequently, finite-lifetime effects on the pulse velocities should be negligible. Finally, we remark that the many soliton solutions~\cite{cau73} to Eq.~\eqref{eq-nonlinear_sigma} are particularly relevant to the observations of Ref.~\onlinecite{pol81a} since the two-soliton solution describes soliton scattering without distortion.

I thank P. W. Anderson, K. DeConde, W. P. Halperin, J. W. Serene, and D. L. Stein for interesting discussions, and especially P. G. N. deVegvar and D. M. Lee for discussions on the experiments of Ref.~\onlinecite{pol81a}. This research was supported by the National Science Foundation under Contract No. DMR-802063.
%
%---------------------------------------------------------------------------------
%\bibliographystyle{apsrev4-1_PRX_style}
%\bibliography{QFS,CM,Books}
%---------------------------------------------------------------------------------
%
%
\end{document}